\documentclass[conference]{IEEEtran}
\usepackage{graphicx}
\usepackage{amsfonts,color}
\usepackage[cmex10]{amsmath}
\usepackage{cite}
\usepackage{amssymb}
\usepackage{epsfig}
\usepackage{latexsym}
\usepackage{epstopdf}
\usepackage{amsmath, amsthm, amssymb}
\usepackage{verbatim}
\usepackage{algorithm}
\usepackage{subcaption}
\usepackage[utf8]{inputenc}
\usepackage{algorithmic}
\usepackage{soul}

\begin{document}
		\title{Mirror-Prox SCA Algorithm for Multicast Beamforming and Antenna Selection}

		\author{\IEEEauthorblockN{Mohamed S. Ibrahim \IEEEauthorrefmark{1}, Aritra Konar \IEEEauthorrefmark{2}, Mingyi Hong \IEEEauthorrefmark{1} and Nicholas D. Sidiropoulos\IEEEauthorrefmark{2}	}
			\IEEEauthorblockA{	\IEEEauthorrefmark{1}Department of Electrical and Computer Engineering, University of Minnesota, Minneapolis, MN\\
				\IEEEauthorrefmark{2}Department of Electrical and Computer Engineering, University of Virginia, Charlottesville, VA\\
				Email: \{\tt youss037@umn.edu, ak8cm@virginia.edu, mhong@umn.edu, nikos@virginia.edu\}
			}
		}
\maketitle

\begin{abstract}
This paper considers the (NP-)hard problem of joint multicast beamforming and antenna selection. Prior work has focused on using Semi-Definite relaxation (SDR) techniques in an attempt to obtain a high quality sub-optimal solution. However, SDR suffers from the drawback of having high computational complexity, as SDR lifts the problem to higher dimensional space, effectively squaring the number of variables. This paper proposes a high performance, low complexity Successive Convex Approximation (SCA) algorithm for max-min SNR ``fair" joint multicast beamforming and antenna selection under a sum power constraint. The proposed approach relies on iteratively approximating the non-convex objective with a series of non-smooth convex subproblems, and then, a first order-based method called Saddle Point Mirror-Prox (SP-MP) is used to compute optimal solutions for each SCA subproblem. Simulations reveal that the SP-MP SCA algorithm provides a higher quality and lower complexity solution compared to the one obtained using SDR.        

\end{abstract}
	
\section{introduction}
Multicast beamforming can be used to optimize the quality of service in shared content (e.g., video) streaming to multiple users, by effectively utilizing the broadcast nature of the wireless medium. In single group multicast, all the subscribers are interested in receiving the same data from the Base Station (BS). Hence, the achievable rate is determined by the minimum received Signal-to-Noise Ratio (SNR) among all the receivers. This renders the problem of maximizing the minimum received SNR subject to a bound on the transmitted power very important~\cite{sidiropoulos2006transmit,konar2017fast}. 

The problem was shown to be NP-Hard in~\cite{sidiropoulos2006transmit}; however, a high quality approximate solution was developed based on the Semi-Definite Relaxation (SDR) technique~\cite{luo2010semidefinite}. On the other hand, the problem of max-min ``fair" single group multicast beamforming subject to per antenna power constraints was studied in~\cite{konar2017fast}. The authors considered a different approximation approach namely Successive Convex Approximation (SCA)~\cite{beck2010sequential}. This technique is based on iteratively solving a sequence of convex problems obtained by constructing a convex surrogate of the original non-convex problem at each iteration. Then each SCA subproblem is solved using a modified version of the Alternating Direction Method of Multipliers algorithm (ADMM). In addition to ADMM, \cite{konar2017fast} also considered the SP-MP method for max-min fair multicasting.

Equipping the BS with more antennas results in a drastic increase in the channel capacity and leads to much more reliable transmission. However, despite the gains that can be achieved, this is accompanied by an increase in hardware complexity. Antenna selection strategies are becoming increasingly desirable as a way to alleviate such complexity. Antenna subset selection has been considered for point-to-point multiple-input multiple-output links~\cite{nai2010beampattern,sanayei2004antenna}, and for receive beamforming~\cite{dua2006receive}. The problem of multicast beamforming with antenna selection was studied in~\cite{mehanna2012multicast}. A SDR-based approach followed by a randomization process was considered to obtain an approximate solution for the problem. However, SDR is computationally expensive as it requires lifting the problem into higher dimensional space, thereby squaring the number of variables. This motivates us to search for an approach which returns a high quality solution, and, at the same time, is much more computationally efficient than convex relaxation.  

This paper aims at solving the problem of joint multicast beamforming and antenna selection in the context of maximizing the minimum received SNR, subject to a sum power constraint. Although the problem is (NP-)hard, SCA is used to approximate the non-convex objective with a class of convex approximations. Then, the Saddle Point Mirror-Prox (SP-MP) algorithm  is utilized to solve each convex subproblem. SP-MP is a first order based method that can be considered as a variant of the mirror descent algorithm. Simulations demonstrate the superior performance of the SP-MP SCA approach over the SDR-based one in terms of both solution quality and run time. 


Regarding notation, matrices (vectors) are denoted by upper- (lower-) case boldface letters, and $(\cdot)^T$, and $(\cdot)^H$ stand for transpose, and conjugate-transpose, respectively. Scalars are represented in the normal face, while calligraphic font is used to denote sets. $ \lVert.\rVert_2 $, $ \lVert.\rVert_1 $, and $  \lVert.\rVert_0  $ denote the $ \ell_2 $-, $ \ell_1 $-, and $ \ell_0 $-norms, respectively. $ \nabla(.) $ denotes the gradient operator. Finally, $ {\bf I}_N $ and $ {\bf 0}_N $ denote the $ N \times N $ identity matrix and the $ N \times 1 $ zero vector, respectively.

\section{Preliminaries} \label{Model}
\subsection{Basic Model}
Consider a multicast scenario consisting of a single base station (BS) with $ N $ antennas and $ M $ single-antenna users. The BS uses an $ N \times 1 $ beamforming weight vector $ \mathbf{w} \in \mathbb{C}^N $ to convey a zero-mean and unit-variance multicast signal $ s $ to all users. The received signal at the $ m $-th user is given by $$ y_m  = {\bf h}_m^H {\bf w} s + z_m, \quad \forall m \in \{1,\cdots,M\}$$
where $ \mathbf{h}_m \in \mathbb{C}^N$ denote the downlink channel between the BS and the $ m $-th receiver. $ z_m $ is zero-mean, wide-sense stationary additive noise at the $ m $-th receiver with variance $ \sigma_m^2 $, and is independent of $ \mathbf{h}_m $ and $ s $. It is further assumed that the $ \mathbf{h}_m $'s  and their corresponding noise variances $ \sigma_m^2 $'s at the users are known at the BS. 

Based on the received signal, the performance of each user can be characterized by the SNR. The received SNR at the $ m $-th user can be written as $ {\bf w}^H {\bf Q}_m {\bf w}$, where $ \mathbf{Q}_m := \frac{\lvert \mathbf{h}_m \mathbf{h}^H_m \rvert^2 }{\sigma^2_m} \succeq \mathbf{0}$, $ \forall \; m \in [M]:=\{1, \cdots, M\} $. Designing the beamformer that maximizes the minimum received SNR among all users under a total BS power constraint, and without performing any antenna selection, can be written as
\begin{subequations}\label{BM1}
	\begin{align}
	& \underset{{\bf w}\in \mathbb{C}^N}{\max} ~ \underset{m \in [M]}{\min} ~ \mathbf{w}^H \mathbf{Q}_m \mathbf{w}\\
	& \text{s.t.} \quad \ \| {\bf w} \|_2^2 \leq P
	\end{align}
\end{subequations}
where $ P $ represents the available power at the BS.
      
\subsection{Antenna Selection}
Suppose now that only a subset of $ K \leq N $ antennas can be active at the BS. The problem of interest is to jointly select the best $ K $ out of $ N $ antennas, and find the corresponding beamforming vector $ \mathbf{w}$ so that the minimum received SNR over all intended receivers is maximized, subject to a bound on the transmitted power. Since the two problems are tightly coupled, the joint problem can be expressed as
\begin{subequations}\label{AS1}
	\begin{align}
	& \underset{{\bf w}\in \mathbb{C}^N}{\max} ~ \underset{m \in [M]}{\min} ~ \mathbf{w}^H \mathbf{Q}_m \mathbf{w}\\
	& \text{s.t.} \quad \ \| {\bf w} \|_2^2 \leq P,\quad \| {\bf w} \|_0 \leq K
	\end{align}
\end{subequations}
where the $ \ell_0 $-norm represents the number of non-zero entries of $ \mathbf{w} $. Since $ \| {\bf w} \|_0 $ is non-convex, the $ \ell_0 $ penalty can be employed to promote sparsity. Then the $ \ell_1 $-norm is typically used as a convex surrogate for the $ \ell_0 $-norm. Thus, (\ref{AS1}) is replaced by
\begin{subequations}\label{AS2}
	\begin{align}
	& \underset{{\bf w}\in \mathbb{C}^N}{\max} ~ \underset{m \in [M]}{\min} ~ \mathbf{w}^H \mathbf{Q}_m \mathbf{w} + \lambda\| {\bf w} \|_1\\
	& \text{s.t.} \quad \ \| {\bf w} \|_2^2 \leq P 
	\end{align}
\end{subequations}
where $ \lambda $ is a positive tuning parameter that adjusts the sparsity of the solution, and thus the  number of selected antennas.
Although (\ref{AS2}) is a non convex problem, SDR can be used to find an approximate solution~\cite{mehanna2012multicast}. In this approach, as the obtained solution is generally not rank 1, a randomization algorithm is used as a final step to extract the beamforming vector. However, the overall procedure can be very computationally demanding owing to the complexity of SDR.  This motivates the need to develop a low-complexity SCA algorithm that can efficiently yield high-quality approximate solutions for (\ref{AS2}).

\subsection{Group-sparsity Inducing Norms}
Regularizing by the $ \ell_1 $-norm (defined as $ \| {\bf w} \|_1 = \sum\limits_{n=1}^{N} |w(n)|$) is known to induce sparsity in the sense that a number of entries of the optimal solution $ {\bf w}^{\star} $, depending on the strength of the regularizer $ \lambda $, will be exactly equal to zero. However, as the proposed approach operates in the real domain,  (see next section), applying the $ \ell_1 $-norm does not imply antenna selection because the zero components of the beamforming vector will not necessarily align to the same antenna. Therefore, it is desired to remove simultaneously the real entry and its corresponding imaginary part in order to switch off a given antenna.

The widely used group-sparsity promoting regularization, which was first introduced in~\cite{yuan2006model}, is the mixed $ \ell_{1,2} $-norm. Such a norm, for instance, may take the form
\begin{equation}\label{L12}
     \|{\bf w} \|_{1,2} :=  \sum\limits_{g}    \|{\bf w}_g\|_2 
\end{equation} 
where each sub-vector $ {\bf w}_g $ is composed of a group of entries selected from the original vector $ \bf w $. It is worth mentioning that the $ \ell_{1,2} $-norm behaves like an $ \ell_1 $-norm on the vector ${\bf w}_g$, and therefore, $ \ell_{1,2} $-norm induces group sparsity. It is obvious that the $ \ell_{1,2} $-norm reduces to the $ \ell_1 $-norm when each group has one entry only.

On the other hand, the dual norm is important to study sparsity-inducing regularization~\cite{jenatton2011structured}. The dual norm $ \|.\|^{\star} $ of the norm $ \|.\| $ is defined for any vector $ {\bf x} \in \mathbb{R}^N$ by
\begin{equation} \label{DN}
       \| {\bf x} \|^{\star} = \underset{\|{\bf s}\| \leq 1}{\max} {\bf s}^T{\bf x}
\end{equation} 
In the next section, we will utilize the definition of the dual norm in addition to the group-sparsity to reformulate problem (\ref{AS2}) such that it can be successively approximated using the SP-MP algorithm.  
\section{Problem Formulation}\label{form}
First, we express (\ref{AS2}) in the real domain. Let $ \bar{{\bf w}} = [{\bf w}_r^T,{\bf w}_i^T]^T \in \mathbb{R}^{2N}$, where $ {\bf w}_r = {\rm I\!Re}\{{\bf w}\} $ and $ {\bf w}_i = {\rm I\!Im}\{{\bf w}\}  $ represent the real and imaginary components of the complex beamforming vector $ {\bf w} $, respectively, and the matrix $ \bar{{\bf Q}}_m \in \mathbb{R}^{2N \times 2N}$ is \\
\begin{equation}
\bar{{\bf Q}}_m = 
\begin{bmatrix} 		
{\rm I\!Re}\{{\bf Q}_m\} & -{\rm I\!Im}\{{\bf Q}_m\} \\
{\rm I\!Im}\{{\bf Q}_m\} & {\rm I\!Re}\{{\bf Q}_m\} 
\end{bmatrix} , \forall \; m \in [M]
\end{equation}
Therefore, (\ref{AS2}) can be equivalently written in terms of real variables as
\begin{subequations}\label{PF1}
	\begin{align}
	& \underset{\bar{{\bf w}}\in \mathbb{R}^{2N}}{\max} ~ \underset{m \in [M]}{\min} ~ \bar{{\bf w}}^H \bar{{\bf Q}}_m \bar{{\bf w}} + \lambda\| \bar{{\bf w}} \|_1\\
	& \text{s.t.} \quad \ \| \bar{{\bf w}} \|_2^2 \leq P 
	\end{align}
\end{subequations}
Since $ \underset{{\bf x} \in \mathcal{X}}{\max} \underset{m \in [M]}{\min} f_m({\bf x}) \Leftrightarrow \underset{{\bf x} \in \mathcal{X}}{\min} \underset{m \in [M]}{\max} -f_m({\bf x})  $~\cite{konar2017fast}, by defining $\tilde{{\bf Q}}_m = - \bar{{\bf Q}}_m $, $ \forall\, m \in [M]$, (\ref{PF1}) can be equivalently expressed as 
\begin{subequations}\label{PF2}
	\begin{align}
	& \underset{\bar{{\bf w}}\in \mathbb{R}^{2N}}{\min} ~ \underset{m \in [M]}{\max} ~ \bar{{\bf w}}^H \tilde{{\bf Q}}_m \bar{{\bf w}} + \lambda\| \bar{{\bf w}} \|_1\\
	& \text{s.t.} \quad \ \| \bar{{\bf w}} \|_2^2 \leq P 
	\end{align}
\end{subequations}
 
Since the problem is written in the real domain, and using the $ \ell_1 $-norm does not guarantee that each of the real and its corresponding imaginary component are zero at the same time, we use the mixed $ \ell_{1,2}$-norm in (\ref{L12}) for group sparsity. Therefore, problem (\ref{PF2}) is modified as 
\begin{subequations}\label{PF3}
	\begin{align}
	& \underset{\bar{{\bf w}}\in \mathbb{R}^{2N}}{\min} ~ \underset{m \in [M]}{\max} ~ \bar{{\bf w}}^H \tilde{{\bf Q}}_m \bar{{\bf w}} + \lambda\| \bar{{\bf w}} \|_{1,2}\\
	& \text{s.t.} \quad \ \| \bar{{\bf w}} \|_2^2 \leq P 
	\end{align}
\end{subequations}
where $\| \bar{{\bf w}} \|_{1,2} =  \sum\limits_{j = 1}^{N}  \|[\bar{{\bf w}}(j),\bar{{\bf w}}(j{+}N)]^T \|_2$, and $ \bar{{\bf w}}(j) $ represents the $ j $-th entry of the vector $ \bar{{\bf w}} $. Since the dual norm of the $ \ell_{1,2} $-norm is the $ \ell_{\infty,2} $-norm~\cite{francis}, using the dual norm definition in (\ref{DN}), problem (\ref{PF3}) can be reformulated as 
\begin{subequations}\label{PF4}
	\begin{align}
	& \underset{\bar{{\bf w}}\in \mathbb{R}^{2N}}{\min} ~ \underset{m \in [M]}{\max} ~ \bar{{\bf w}}^H \tilde{{\bf Q}}_m \bar{{\bf w}} + 
	\lambda \underset{\| {\bf s}\|_{\infty,2} \leq 1}{\max} {\bf s}^T\bar{{\bf w}}\\
	& \text{s.t.} \quad \ \| \bar{{\bf w}} \|_2^2 \leq P 
	\end{align}
\end{subequations} 
Note that the problem remains non convex, as the point-wise maximum of concave functions is not convex. Therefore, we propose an iterative SCA approach to approximate the original non-convex problem by a sequence of convex problems. By defining $ g(\bar{{\bf w}}):= \underset{m \in [M]}{\max} ~ \bar{{\bf w}}^H \tilde{{\bf Q}}_m \bar{{\bf w}} $, it can be replaced by its piecewise-linear approximation about $ \bar{{\bf w}}^{(n)} $~\cite{konar2017fast}. Let $ {\bf a}_m^{(n)} = 2\tilde{{\bf Q}}_m\bar{{\bf w}}^{(n)} \in \mathbb{R}^{2N} $ and  $ b_m^{(n)} = -\bar{{\bf w}}^{(n)T}\tilde{{\bf Q}}_m\bar{{\bf w}}^{(n)} \in \mathbb{R}$,  where $ n $ denotes the $ n $-th iteration. Thus, problem (\ref{PF4}) can be written as
\begin{subequations}\label{PF5}
	\begin{align}
	& \underset{\bar{{\bf w}}\in \mathbb{R}^{2N}}{\min} ~ \underset{m \in [M]}{\max} ~ {\bf a}_m^{(n)T}\bar{{\bf w}}+b_m^{(n)} + 
	\lambda \underset{\| {\bf s}\|_{\infty,2} \leq 1}{\max} {\bf s}^T\bar{{\bf w}}\\
	& \text{s.t.} \quad \ \| \bar{{\bf w}} \|_2^2 \leq P 
	\end{align}
\end{subequations}
It is obvious that (\ref{PF5}) is a convex problem, as the maximization of a piece-wise affine function is a convex optimization problem. We now define the matrix $ {\bf A}^{(n)} = [{\bf a}_1^{(n)}, \cdots, {\bf a}_M^{(n)}]^T  \in \mathbb{R}^{M \times 2N}$ and the vector $ {\bf b}^{(n)} = [b_1{(n)}, \cdots, b_M{(n)}]^T  \in \mathbb{R}^{M}$. Note that the maximization of a piece-wise linear function is equivalent to maximizing a linear function over the $ M $-dimensional probability simplex, with the maximum attained at one of the vertices (i.e., a canonical basis vector of $\mathbb{R}^{M}$). More formally, (\ref{PF5}) can be written as
\begin{subequations}\label{PF6}
	\begin{align}
	& \underset{\bar{{\bf w}}\in \mathbb{R}^{2N}}{\min} ~ \underset{{\bf y}\in \bigtriangleup_M}{\max} ~ {\bf y}^T({\bf A}_m^{(n)}\bar{{\bf w}}+{\bf b}_m^{(n)}) + 
	\lambda \underset{\| {\bf s}\|_{\infty,2} \leq 1}{\max} {\bf s}^T\bar{{\bf w}}\\
	& \text{s.t.} \quad \ \| \bar{{\bf w}} \|_2^2 \leq P 
	\end{align}
\end{subequations}    
where $ \bigtriangleup_M {:=} \big\{{\bf y} \in \mathbb{R}_{+}^{M} ~|~ \sum\limits_{i=1}^{M} y_i = 1\big\}$. Also, we define the sets, $ \mathcal{S} := \big\{{\bf s} \in \mathbb{R}^{2N} ~|~ \underset{i}\max ~ \|[ s(i), s(i{+}N)]^T \|_2 \leq 1 \big\} $ and $ \mathcal{W} := \big\{ \bar{\bf w} \in \mathbb{R}^{2N} ~|~  \| \bar{{\bf w}} \|_2^2 \leq P \big\} $. Note that the objective of (\ref{PF6}) is a bi-linear function, which is convex in $ \bar{\bf w} $ for given $ (\bf y,\bf s) $ and concave in $ (\bf y,\bf s)$ for fixed $\bar{\bf w} $. Therefore, as a final step, let us write ($ \ref{PF6} $) in more compact form by defining the matrix $ \bar{\bf A} = [{\bf A}^T,\lambda{\bf I}_{2N}]^T$, the vector $ {\bar{\bf x}} = [{\bf y}^T, {\bf s}^T]^T $ and the vector $ \bar{\bf b} = [{\bf b}^T, {\bf 0}_{2N}^T]^T$. Hence, ($ \ref{PF6} $) can be equivalently reformulated as
\begin{equation}\label{PF7}
       \underset{\bar{{\bf w}} \in \mathcal{W}  }{\min} ~ \underset{\bar{\bf x}\in \bigtriangleup_M \times \mathcal{S}}{\max} ~ \bar{\bf x}^T(\bar{\bf A}_m^{(n)}\bar{{\bf w}}+\bar{\bf b}^{(n)}) 
\end{equation}
Defining $ \phi^{(n)}(\bar{\bf w},\bar{\bf x}) := \bar{\bf x}^T(\bar{\bf A}_m^{(n)}\bar{{\bf w}}+\bar{\bf b}^{(n)}) $ and the set $ \mathcal{X} :=  \bigtriangleup_M \times \mathcal{S} $,  ($ \ref{PF7} $) can be expressed as
\begin{equation}\label{PF8}
\underset{\bar{{\bf w}} \in \mathcal{W}  }{\min} ~ \underset{\bar{\bf x}\in \mathcal{X}}{\max} ~ \phi^{(n)}(\bar{\bf w},\bar{\bf x}) 
\end{equation}
Since $\phi^{(n)}(.,.)$ is bilinear and $\mathcal{X}$ and $\mathcal{W}$ are both simple, convex, compact sets, by Sion's Minimax theorem, we have
\begin{equation}
\underset{\bar{{\bf w}} \in \mathcal{W}  }{\min} ~ \underset{\bar{\bf x}\in \mathcal{X}}{\max} ~ \phi^{(n)}(\bar{\bf w},\bar{\bf x})  = \underset{\bar{{\bf x}} \in \mathcal{X}  }{\min} ~ \underset{\bar{\bf w}\in \mathcal{W}}{\max} ~ \phi^{(n)}(\bar{\bf w},\bar{\bf x}) 
\end{equation}
which implies that the optimal solution pair $(\bar{\bf w}^*,\bar{\bf x}^*)$ of \eqref{PF8} is a saddle-point of $\phi^{(n)}(\bar{\bf w},\bar{\bf x})$.
In the next section, we describe how SP-MP can be used to efficiently compute $(\bar{\bf w}^*,\bar{\bf x}^*)$.  
\section{SP-MP for SCA}\label{SP-MP}
In~\cite{Nemirovski}, Nemirovski devised a simple prox-type method to solve the saddle-point problem $ \underset{{\bf x} \in \mathcal{X}}\min~\underset{{\bf y} \in \mathcal{Y}}\max ~ \phi(\bf x,y) $ contingent on the sets $\mathcal{X}$ and $\mathcal{Y}$ being ``simple enough'' convex compact sets. This method is known as SP-MP, which can be considered as a variant of the mirror descent algorithm~\cite{beck2003mirror}. We now briefly outline the SP-MP algorithm that is used to solve (\ref{PF8}).

First, let $ \phi_{\mathcal{X}} (\bf x)$, $ \phi_{\mathcal{W}} (\bf w)$  denote ``mirror maps'' for the sets $\mathcal{X}$ and $\mathcal{W}$ (i.e., strongly convex functions capable of exploiting the geometry of the sets), respectively. Also,  in mirror descent, the projection is done via the Bregman divergence associated to $ \phi $. This can be defined as
\begin{equation} \label{BD}
  D_{\phi}(\bf x,{\bf x'}) = \phi(\bf x) - \phi({\bf x'}) - \nabla\phi({\bf x'})^{T} (\bf x - {\bf x'})
\end{equation}
The function $ \phi(\bf w, \bf x) $ is said to be $ (\beta_{11},\beta_{12},\beta_{21},\beta_{22}) $-smooth if for any $ \bf x, \bf x'  \in \mathcal{X}$ and   
$ \bf w, \bf w'  \in \mathcal{W}$,
\begin{subequations}\label{MP2}
	\begin{align}
	 &\| \nabla_{\bf w}  \phi(\bf w, \bf x) - \nabla_{\bf w}  \phi(\bf w', \bf x) \|^{\star}_{\mathcal{W}} \leq \beta_{11} \|\bf w - \bf w'\|_{\mathcal{W}},\\
	 &\| \nabla_{\bf x}  \phi(\bf w, \bf x) - \nabla_{\bf x}  \phi(\bf w, \bf x') \|^{\star}_{\mathcal{X}} \leq \beta_{22} \|\bf x - \bf x'\|_{\mathcal{X}}, \\
	 &\| \nabla_{\bf w}  \phi(\bf w, \bf x) - \nabla_{\bf w}  \phi(\bf w, \bf x') \|^{\star}_{\mathcal{W}} \leq \beta_{12} \|\bf x - \bf x'\|_{\mathcal{X}}, \\
	 &\| \nabla_{\bf x}  \phi(\bf w, \bf x) - \nabla_{\bf x}  \phi(\bf w', \bf x) \|^{\star}_{\mathcal{X}} \leq \beta_{21} \|\bf w - \bf w'\|_{\mathcal{W}} 
	\end{align}
\end{subequations}
where  $ \|.\|^{\star}_{\mathcal{W}}$ and $ \|.\|^{\star}_{\mathcal{X}}$ indicate the dual norms of $ \|.\|^{\star}_{\mathcal{W}}$ and $ \|.\|^{\star}_{\mathcal{W}}$, respectively.
We now consider the mirror map $ \phi(\bf z) = \phi(\bf w, \bf x) = \phi_{\mathcal{X}} (\bf x) + \phi_{\mathcal{W}} (\bf w)$, where $ \mathcal{Z} = \mathcal{X} \times \mathcal{W} $ and  $ \mathcal{X} = \mathcal{Y} \times \mathcal{\bigtriangleup}_M $. Furthermore, defining $ \beta := \max\{\beta_{ij}\} $, for $ i,j \in \{1,2\} $, and the step size $ \alpha = \frac{1}{2\beta} $, the SP-MP algorithm is given by the following steps
\begin{algorithm} \label{A1}
	\caption{Saddle Point Mirror-Prox}
\textbf{initialization}: Define $ {\bf z}_t = [\bar{{\bf w}}_t^T,\bar{{\bf x}}_t^T] $, 
$ {\bf r} = [\bar{{\bf u}}_t^T,\bar{{\bf v}}_t^T] $, 
$ \psi({{\bf z}}_t)  = [\nabla_{\bar{{\bf w}}} \phi(\bar{{\bf w}}_t,\bar{{\bf x}}_t)^T,-\nabla_{\bar{{\bf x}}} \phi(\bar{{\bf w}}_t,\bar{{\bf x}}_t)^T$], and
$\psi(\bar{{\bf r}})  = [\nabla_{\bar{{\bf u}}} \phi(\bar{{\bf u}}_t,\bar{{\bf v}}_t)^T,-\nabla_{\bf v} \phi(\bar{{\bf u}}_t,\bar{{\bf v}}_t)^T$] for $ t \geq 0 $, starting with feasible $ \bar{{\bf z}}_0 $   \\ 
\textbf{Repeat:}
	\begin{enumerate}
\item $\nabla \phi({\bf r}'_{t+1}) = \nabla \phi({{\bf z}}_t) - \alpha \psi({{\bf z}}_{t}) $
\item $ {\bf r}'_{t+1} = \nabla \phi^{-1}(\nabla \phi({{\bf z}}_t) - \alpha \psi({{\bf z}}_{t}))$
\item $  {\bf r}_{t+1}  = \arg \min_{{\bf z} \in \mathcal{Z}} D_{\phi}({{\bf z}},{\bf r}'_{t+1})$
\item $\nabla \phi({\bf z}'_{t+1}) = \nabla \phi({{\bf z}}_t) - \alpha \psi({\bf r}_{t+1}) $
\item $ {\bf z}^{'}_{t+1} = \nabla \phi^{-1}(\nabla \phi({{\bf z}}_t) - \alpha \psi({{\bf r}}_{t+1}))$
\item $ {\bf {z}}_{t+1}  = \arg \min_{{\bf {z}} \in \mathcal{Z}} D_{\phi}({{\bf z}},{{\bf z}'}_{t+1})$
\item set $t = t+1$
	\end{enumerate}
		\textbf{Until} termination criterion is met
\end{algorithm}

Define the mirror map for the sets $ \mathcal{W}, \mathcal{\bigtriangleup}_M ~ \text{and} ~ \mathcal{S}  $ to be $ \phi(\bar{\bf w}) = \| \bar{{\bf w}} \|_2^2  $, $  \phi({\bf y}) = \sum\limits_{m=1}^{M} { y}_m \log {y}_m $ and $ \phi({\bf s}) = \| s \|_2^2 $, respectively. Thus the mirror map $ \phi({\bf z})  $ defined for $ {\bf z} = (\bar{\bf w},\bar{\bf x}) $ is given by
\begin{eqnarray}\label{MM}
        \phi({\bf z}) &=& \phi(\bar{\bf w}) + \phi(\bar{{\bf x}}) \nonumber \\
                           &=&  \phi(\bar{\bf w}) + \phi({\bf y})+\phi({\bf s}) \nonumber \\
                           &=&  \| \bar{{\bf w}} \|_2^2 + \sum\limits_{m=1}^{M} { y}_m \log{ y}_m + \| {\bf s} \|_2^2 
\end{eqnarray}
Then $ \nabla \phi({\bf z}) $  and $ \nabla^{-1} \phi({\bf z}) $ can easily be obtained as follows
\begin{eqnarray} \label{MD}
         \nabla \phi({\bf z}) &{=}& [\bar{\bf w},1+\log{y}_1,\cdots,1+\log{y}_M,{\bf s}] \\
\nabla^{-1} \phi({\bf z}) &{=}& [\bar{\bf w},\exp({y}_1 -1),\cdots,\exp({y}_M -1),{\bf s}]
\end{eqnarray}
Moreover, by using the definition of the Bregman Divergence in (\ref{BD}), the non-Euclidean projection problem in Algorithm 1 can be written as
\begin{eqnarray}\label{PBD}
\underset{\bar{{\bf z}} \in \mathcal{Z}\cap\mathcal{D}} \min D_{\phi}(\bar{{\bf z}},\bar{{\bf z}}') {=} 
\min_{\substack{\bar{{\bf w}} \in \mathcal{W}\\\bar{{\bf y}} \in \bigtriangleup_M \\ {{\bf s}} \in \mathcal{S}}} \frac{1}{2}\| \bar{{\bf w}} - \bar{{\bf w}}' \|_2^2 +  \frac{1}{2}\| {{\bf s}} - {{\bf s}}' \|_2^2  \nonumber \\
{+}\sum\limits_{m=1}^{M} { y}_m \log\frac{{ y}_m}{{y}'_m} - \sum\limits_{m=1}^{M}({y}_m - {y}'_m) 
\end{eqnarray}
The above problem can be resolved into three distinct projection problems as follows
\begin{subequations}\label{3PP}
	\begin{align}
	\underset{\bar{{\bf w}} \in \mathcal{W}}\min ~ &\frac{1}{2}\| \bar{{\bf w}} - \bar{{\bf w}}' \|_2^2, \\
	\underset{{{\bf s}} \in \mathcal{S}}\min ~ &\frac{1}{2}\| {{\bf s}} - {{\bf s}}' \|_2^2, \\
    \underset{{{\bf y}} \in \bigtriangleup_M}\min ~ &\sum\limits_{m=1}^{M} { y}_m \log\frac{{ y}_m}{{y}'_m} - \sum\limits_{m=1}^{M}({y}_m - {y}'_m)
		\end{align}
\end{subequations} 
Note that problems (\ref{3PP}a) and (\ref{3PP}b) are Euclidean projection onto $ \mathcal{W} $ and $ \mathcal{S} $, respectively. It is easy to show that the projection on the set $ \mathcal{S}$ can be  performed by projecting separately each sub-vector $ {{\bf s}}_j=  [s(j), s(j+N)]^T $ on a unit $ l_2$-ball, where $ j \in \{1,\cdots,N\} $. On the other hand, the projection on the $ M $-dimensional probability simplex has a simple closed form solution~\cite{bubeck} given by 
\begin{equation}
{ {\bf y}} = \left\{\begin{array}{lr}
{ {\bf y}}', & {{\bf y}}' \in \bigtriangleup_M \\
\frac{ { {\bf y}}'}{\lVert  {{\bf y}}^{'} \rVert_1}, &  \text{otherwise}
\end{array}\right\}
\end{equation} 
Finally, the step size $ \alpha = \frac{1}{2L} $ can be obtained from (\ref{MP2}) by noticing that $ \beta_{11} = \beta_{22} =0 $ and $ \beta_{12} = \beta_{21} = L $, where the Lipschitz constant  $ L $  is given by
\begin{equation}
       L = \max(\underset{m}\max(\|{\bf a}^{(n)}_m\|_2),\lambda)
\end{equation}
The steps of the SP-MP algorithm can now be applied to solve (\ref{PF8}) to obtain the optimal solution 
$ \bar{{\bf w}}^{\star(n)} $ at the $ n $-th iteration. Then SCA iterative algorithm is used to get the final solution $ \bar{{\bf w}}^{\star(n)} $ that corresponds to the optimal beam-forming vector. The final solution is sparse where the degree of sparsity depends on the strength of the regularizer $ \lambda $. A binary search method is used to get the optimal $ \lambda $ that yields the desired set of antennas required for transmission. The procedure of the binary search method can be described as follows. For a given upper bound $ \lambda_{UB} $ and lower bound $ \lambda_{LB} $, we set $ \lambda = \frac{\lambda_{UB} - \lambda_{LB}}{2} + \lambda_{LB}$ and run the MP algorithm. Let $ S $ denote the number of non zero entries in the beamforming vector. If $ S = K $, then we are essentially done, just run the MP algorithm one more time with the reduced number of antennas to obtain the optimal beamforming vector. Otherwise, if $ S > K $, set $\lambda_{LB} = \lambda  $, while if $ S< K $, set $ \lambda_{UB} = \lambda $, and repeat until $ S = K $. Now, the overall algorithm is given by
\begin{algorithm} \label{A2}
	\caption{SP-MP SCA}
	\textbf{initialization}: Randomly generate a feasible starting point $ \bar{{\bf w}}^{(0)} \in \mathcal{W} $, set $ n:=0 $ \\
	\quad \textbf{Repeat:}
	  \begin{itemize}
	  	\item compute $  \bar{{\bf w}}^{(n+1)} $ using the SP-MP algorithm.
	  	\item set $ n:=n+1 $
	  \end{itemize}
	\textbf{Until} termination criterion is met 

\end{algorithm}

\section{Experimental Results}\label{Simu}
To test the performance of the SP-MP algorithm, we consider a scenario where a BS with $ N = 30 $ antennas broadcasts a common message to $ M = 50 $ receivers. The noise variance was set to 1, and the total transmission power P was set to 10. The SCA algorithm was run for 10 iterations, with 1000 iterations being used to solve each SCA subproblem using the SP-MP algorithm. The SP-MP SCA was implemented in MATLAB. Moreover, it was empirically found that, setting $ \lambda_{UB} = 2 $ and $ \lambda_{LB} = 0 $ is sufficient to cover the required range of $ \lambda $ for the binary search to get the optimal $ \lambda $ required for the desired sparse solution. The results were averaged over 200 Monte-Carlo trials. The downlink channels $ \{{\bf h}_m^H\}_{m=1}^{M} $ are modeled as 
                    \begin{equation}
                     {\bf h}_m^H = \sqrt{\frac{N}{L_m}} \sum\limits_{l=1}^{L_m} \alpha_m^{(l)}{\bf a}_t(\theta^{(l)})^H, \forall m=1,\cdots,M
                    \end{equation}
where $ L_m \sim \mathcal{U}[4,5,\cdots,10]$ is the number of scattering paths between the BS and the $ m $-th user, $ \alpha_m^{(l)} \sim \mathcal{CN}(0,1)$ is the complex gain of the $ l $-th path, $ {\bf a}_t(.) $ is the array response vector at the transmitter, and $ \theta^{(l)} \sim \mathcal{U}[-\pi/2,\pi/2]$ denotes the azimuth angle of departure of the $ l $-th path. Assuming the BS is equipped with a uniform linear array, then
                \begin{equation}
                {\bf a}_t(\theta) = [1,\exp^{ikd\sin(\theta)}, \cdots,\exp^{ikd(N-1)\sin(\theta)}]
               \end{equation}
where $ k = 2\pi/\lambda$, $ \lambda $ is the carrier wavelength and $ d = \lambda/2$ is the spacing between antenna elements.               

To compute the optimal solution of (\ref{AS2}), we run exhaustive search and use it as a performance benchmark. Furthermore, to show the effectiveness of SP-MP algorithm in yielding a high-quality solution, the SDR-based algorithm~\cite{mehanna2012multicast} was included in comparison. The modeling language YALMIP which is implemented as a free toolbox in MATLAB~\cite{yal}, and uses SeDuMi ~\cite{sedumi} for the actual computations, was used to generate the SDR-based solution of (\ref{AS2}). All experiments were carried out on a Linux Machine with 8 Intel i7 cores and 8 GB of RAM.

\begin{figure}
    	\centering
    	\includegraphics[width=70mm]{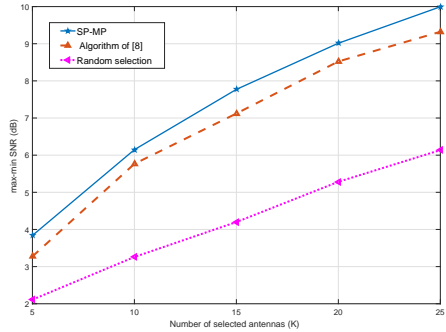}
    	\caption{max-min SNR vs $ K $ with $ N = 30 $ and $ M=50 $}
    	\label{fig:SNR}
    \end{figure}

\begin{figure}
     	\centering
     	\includegraphics[width=70mm]{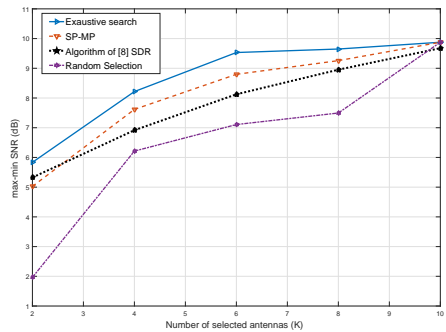}
     	\caption{max-min SNR vs $ K $ with $ N = 10 $ and $ M=16 $}
     	\label{fig:SNR2}
     \end{figure}

\begin{figure}
  	\centering
  	\includegraphics[width=70mm]{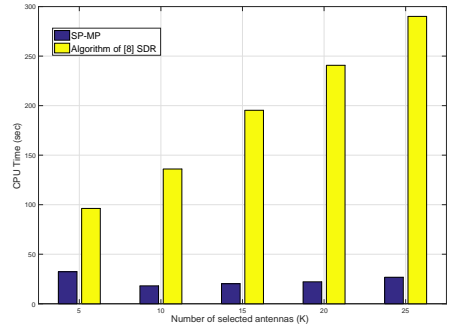}
  	\caption{CPU Time vs $ K $ with $ N = 30 $ and $ M=50 $}
  	\label{fig:time}
  \end{figure}

The performance with respect to the max-min SNR attained is shown in Figures~\ref{fig:SNR},\ref{fig:SNR2} for different values of $ N $ and $ M $, while timing results are depicted in Fig~\ref{fig:time}. It is obvious that, for $ N = 30 $ the SP-MP algorithm returns a high quality solution compared to that of SDR for all values $ K $. While, for $ N = 10 $, the max-min SNR achieved using the algorithm of~\cite{mehanna2012multicast} is slightly better than that of SP-MP, at $ K = 2 $ only, with about 0.3 dB. Note that exhaustive search was not included in Fig~\ref{fig:SNR}, as we can not afford running it when $ N = 30 $.  Finally, Fig~\ref{fig:time} shows that the SP-MP algorithm is much cheaper than SDR-based algorithm. In particular, SP-MP is up to approximately 10 times faster than SDR-based algorithm.


\section{Conclusions} \label{Conc} 
The (NP-)hard problem of single-group multicast beamforming with antenna selection was studied. The goal is to jointly select the optimal set of antennas and their corresponding beamforming vector such that the minimum received SNR at the intended receivers is maximized. The problem was reformulated to be successively approximated using a first order-based method namely SP-MP. First, a SCA approach was adopted to obtain approximate solutions by iteratively solving a sequence of convex approximation of the original non-convex problem. Then, each subproblem was solved using SP-MP which can be considered as a variant of the mirror descent algorithm. Simulations revealed that the SP-MP was better in terms of both solution quality and run time.	
\bibliographystyle{IEEEtran}
\bibliography{IEEEabrv,refrences}

\end{document}